\documentstyle[aps,bezier,epsfig,amsmath,amssymb]{revtex}

\begin{document}

\newcommand{\be}{\begin{eqnarray}}
\newcommand{\ee}{\end{eqnarray}}
\newcommand{\beq}{\begin{equation}}
\newcommand{\eeq}{\end{equation}}
\newcommand{\xx}{\begin{eqnarray*}}
\newcommand{\yy}{\end{eqnarray*}}
\newcommand{\nn}{\nonumber}
\newcommand{\Vol}{{\rm Vol}}
\newcommand{\sign}{{\rm sign}}
\newcommand{\tr}{{\rm Tr}}
\def\cut{{}\hfill\cr \hfill{}}

\twocolumn[\hsize\textwidth\columnwidth\hsize\csname@twocolumnfalse\endcsname

\title{ Universal Magnetic Fluctuations with a Field Induced Length
Scale}

\author{ B. Portelli$^{1}$,  P.C.W.
Holdsworth$^{1}$, M. Sellitto$^{1}$ and S.T. Bramwell$^{2}$}

\address{
$^1$ Laboratoire de Physique, Ecole Normale Sup\'erieure,
46 All\'ee d'Italie, F-69364 Lyon cedex 07, France. \\
$^2$ Department of Chemistry, University College London, 20
Gordon
Street, London, WC1H~0AJ, United Kingdom.}

\maketitle

\begin{abstract}

We calculate the probability density function for the order parameter
fluctuations in the low temperature phase of the 2D-XY model of
magnetism near the line of critical points. A finite correlation
length, $\xi$, is introduced with a small magnetic field, $h$, and
an expression for $\xi(h)$ is developed
by treating non-linear contributions to the field energy using a Hartree
approximation. We find analytically a series of universal non-Gaussian
distributions of the finite size scaling form
$P(m,L,\xi)\sim L^{\beta/\nu}P_L(mL^{\beta/\nu}, \xi/L)$ and present
a function of the form $P(x) \sim (\exp (x - \exp (x)))^{a(h)}$
that gives the PDF to an excellent approximation. We propose $a(h)$
as an indirect measure of the length scale of correlations in a wide
range of complex systems.
\medskip

\medskip

\noindent{PACS numbers: 05.40.-a, 05.50.+q, 75.10.Hk}

\medskip

\medskip 

\noindent To appear in {\it Phys. Rev. E}

\end{abstract}

\twocolumn\vskip.5pc]\narrowtext


\section{introduction}

There has recently been considerable interest in the fluctuations
of a spatially averaged quantity in systems with
correlations over a macroscopic length scale
$\xi$~\cite{LPF,zol,BHP,har.99,PHL,aum.00,cha.00,aji.01}.
The most accessible example from both an experimental and a
theoretical point of view is a critical system, where the divergence
of the correlation length $\xi$ is interrupted by
the system size $L$.
Here, the breakdown of Landau theory
and the occurrence of a non-analytic fixed point for the free
energy in a renormalization group flow, tells one that the
fluctuations will not
be Gaussian.
Even so, extending the scaling hypothesis to include the macroscopic scale
$L$, one can deduce that the probability density function (PDF)
for order parameter fluctuations should have universal properties. Further,
the PDF must be a homogeneous function of  $m$, $L$ and
$\xi$ of the following form~\cite{bin}
\be \label{scale}
P(m,L,\xi)\sim L^{\beta/\nu} \, P_L(mL^{\beta/\nu}, \,\xi/L) \,.
\ee
Here, adopting the language of a magnetic phase transition,
$\nu$ and $\beta$ are the usual critical exponents relating to
the divergence of the correlation length and the singularity in
the magnetization, $m$.
The scaling hypothesis therefore predicts fluctuations of a universal
form, independently of system size, for constant ratio $\xi/L$.

We study
the low temperature phase of the 2D-XY model, defined by the Hamiltonian
\be\label{eq-H0}
H =-J\sum_{\langle  i,j  \rangle }\cos(\theta_i-\theta_j) -
h \sum_i \cos \theta_i \, .
\ee
The exchange interaction and magnetic field are of strength $J$ and
$h$ and the angle $\theta_i$ gives the orientation of a classical spin vector
of unit length, confined to a plane. We define the  magnetization for a
single configuration
\be\label{eq-mag}
m = \frac{1}{N} \sum_{i=1,N} \cos(\theta_i - \overline{\theta}) \,,
\ee
where $\overline{\theta} = \tan^{-1}\left(\sum_i \sin \theta_i /\sum_i
  \cos \theta_i\right)$ is the instantaneous magnetization
direction.

This is perhaps the simplest
nontrivial system in which one can study critical phenomena.
At low-temperature and in zero field there is a line of
critical points, separated from the high temperature
paramagnetic phase by the Kosterlitz-Thouless-Berezinskii phase
transition. The physics of this low temperature phase is perfectly
captured by a harmonic, or spin wave Hamiltonian. That is,
one can, without loss of generality, develop the cosine interaction
to order $(\theta_i-\theta_j)^2$ and neglect the periodicity of
$\theta_i$. This Hamiltonian is diagonal in reciprocal space and can
be solved straightforwardly.  As a result,
all critical phenomena can be calculated microscopically from Gaussian
integration, without
the need for either the scaling hypothesis, or the renormalization
group. Along the line of critical points the exponents $\beta$ and $\nu$
are not individually defined, but there ratio is and the system
has a single independent exponent $\eta = 2\beta/\nu = T/2\pi J$.

We have previously been interested in the zero field, or
strongly correlated regime where the divergence
of $\xi$ is completely removed by the system size $L$ and $P_L$ becomes a
function of a
single variable $mL^{\beta/\nu}$~\cite{pre}. We have found that 
$P_L$, when plotted as a function of $\mu =(m-\langle m \rangle)/\sigma$, is
a universal function, not only of system size, but also of temperature
and therefore of critical exponent $\eta$. 
Here, $\langle m\rangle$ is the mean and $\sigma$ the standard deviation of the distribution. This rather surprising result
gives weight to our conjecture~\cite{BHP} that the critical 
fluctuations of systems
in certain universality classes are captured, at least qualitatively
by the fluctuations of the 2D-XY model.

In this paper we generalize our previous results by
introducing a second length into the problem with
the aid of a
magnetic field. The field breaks the symmetry moving the system
into an ordered magnetic state with finite correlation length $\xi$.
However, taking a van Hove type thermodynamic limit, with the ratio
$\xi/L$ constant, should lead to a family of limit functions, all with
divergent correlation length, varying in form from
the anisotropic limit (see figure 1) to a Gaussian function, as
the ratio $\xi/L$ falls to zero. In the next section we develop a starting
Hamiltonian that satisfies the requirements of the scaling hypothesis, using
a self-consistent Hartree approximation. In section 3 we give theoretical 
results for the PDF for finite field and compare our results with Monte Carlo
simulation. In section 4 we fit the curve with a generalized form
of Gumbel's first asymptote from extremal statistics.

\section{Hartree approximation and the hyperscaling relation}

Expanding the field energy in small angles, in the same way
as the exchange term
and Fourier transforming gives a convenient starting Hamiltonian
\be\label{eq-Hspin}
H= H_0+ \frac{J}{2} \sum_{\vec q\ne 0}\left( \gamma_q + \frac{h}{J}\right)
\phi_q^2 \,,
\ee
where
$\phi_{q} = Re\left[ 1/\sqrt{ N} \sum_i \theta_i \exp(-i\vec q.
\vec r_i)\right]$
is the real part of the Fourier transformed spin
variable. $\gamma_q = 4 - 2 \cos(q_x) -2 \cos(q_y)$ and
the sum $\sum_{\vec q\ne 0}$ is over the Brillouin zone for a square
lattice with periodic boundaries, with $\vec q$ taking on discrete values
$q_x=(2\pi/L) n_x,\;q_y=(2\pi/L) n_y$, $n_x,n_y = 0,2...\sqrt(N)$.
Here and throughout the paper we have set the nearest neighbour distance
on the lattice equal to unity.
Expanding $\gamma_q$ for small $q$ we can write  the
Green's function propagator
\be\label{eq-G}
G(q)\approx \frac{1}{q^2 + \xi^{-2}},\;\;\; \xi = \sqrt{J/h} \,;
\ee
the magnetic field indeed introduces a length scale $\xi$.

However, this naive starting point needs some development before
proceeding with the calculation as, as it stands it does not satisfy the
well known hyperscaling relation. 
To see this, consider the following scaling
argument~\cite{ber.71}: at the critical temperature but in finite
field,  the thermally averaged magnetization can be
expressed in terms of both $\xi$ and $h$
\be\label{s-rel}
\langle m \rangle  \sim \xi^{-\beta/\nu} \sim 
\left(\frac{h}{J}\right)^{1/\delta} \,,
\ee
where $\delta$ is the usual critical exponent for the response in
finite field. Putting the expression (\ref{eq-G}) for $\xi$ in (\ref{s-rel})
leads to a
relation between the exponents:  $\delta = 2\nu/\beta$, in disagreement
with the hyperscaling relation $\delta + 1 = d\nu/\beta$, which
should be valid for the 2D-XY model~\cite{weak.scaling}. The
 error comes from the development of the field term in
small angles. Even at low temperature, when the nearest neighbour
differences $\theta_i - \theta_j$ are small,
the deviations of $\theta_i$ from the fixed field direction
are divergent in the thermodynamic limit.
The development of the field term in small angles is
therefore invalid. This problem can be dealt with
in the low temperature phase, in the absence of
vortices,  using
the Hartree approximation introduced by
Pokrovsky and Uimin~\cite{Uimin}. 
Expanding $\cos(\theta_i)$ in powers of $\theta_{i}^{2}$
%
%
%
we make a mean field decoupling
\begin{equation}\label{dec}
{\theta_{i}}^{2p}\longrightarrow{\cal C}_{p}\langle {\theta_{i}}^{2p-2}
\rangle {\theta_{i}}^{2}
\end{equation}
 where ${\cal C}_{p}=\frac{(2p)!}{2(2p-2)!}$ is a binomial counting
factor. As the underlying Hamiltonian (\ref{eq-Hspin}) is quadratic
we can reduce $\langle {\theta_{i}}^{2p-2} \rangle $ using Wick's theorem
%
%
and after some resummation we eventually find
\be
\cos \theta_i \approx 1 - \langle m \rangle  \frac{\theta_i^2}{2} \,.
\ee
The field term in the Hamiltonian (\ref{eq-Hspin}) then becomes
\be\label{H-1}
h \sum_i \cos(\theta_i)& =& Nh -
\frac{1}{2} h_{eff}(T) \sum_{\vec q\ne 0} \phi_q^2
\nonumber \\
h_{eff}(T) &=& \langle m \rangle h \, .
\ee
Using the scaling relation 
$\langle m \rangle \sim (h/J)^{\frac{1}{\delta}}$ 
the effective field $h_{eff} \rightarrow h^{{\delta +1\over{\delta}}}$
and the above scaling argument correctly yields the hyperscaling scaling
relationship defined above. Note however, that the scaling argument is valid in
the thermodynamic limit where $\xi/L \ll 1$ and the influence of the 
finite system size is negligible. In the crossover region that interests us,
with $\xi/L \sim O(1)$ one cannot make this substitution and in general 
one must explicitly work with expression (\ref{H-1}).

The point of principle that poses the problem for the hyperscaling 
relation is that making the  substitution
$\cos(\theta_i)\approx 1-(1/2) \theta_i^2$ results in an
order parameter conjugate to the field, $\langle m \rangle  = \langle
1-(1/2) \theta_i^2 \rangle  =
1- T/8\pi J \log (N)$ which diverges
with system size, for any finite temperature. In
 order for the hyperscaling
relation to hold, $\langle m \rangle $ must
be a correctly defined intensive variable. For this to be so, the higher
order terms in the expansion of $\cos \theta_i$ must be retained, at
least within the level of approximation shown here. 
The calculation of
Berezinskii \cite{ber.71} is consistent with this thermodynamic argument.

Still in the absence of vortices, an effective coupling constant $J_{eff}$ can
be calculated in a similar manner. In zero field we
did not need to take account of this, as $P_L$ for
a quadratic Hamiltonian is
independent of temperature throughout the low temperature
regime. However, as the correlation length depends on both the field
and the coupling constant, we now have to calculate it
if we are to have good agreement with numerical data. 
Expanding $\cos(\theta_i-\theta_j)$ in powers of
%
%
the discrete difference operator
$\vec\nabla\theta_{i}$~\cite{ber.71} and
again using the decoupling (\ref{dec}) for $(\vec\nabla\theta_{i})^{2p}$,
 we arrive at a self-consistent expression
for $J_{eff}(T)$~\cite{Uimin,Leoncini}
\begin{equation}\label{H-2}
J_{eff}=J\exp\Bigl(-\frac{T}{4J_{eff}}\Bigr) \,.
\end{equation}
The
Green's function we finally use for the calculation of the PDF is therefore
of the form (\ref{eq-G}) but with the correlation length given in terms
of the self-consistent effective field and coupling constant
\be
 \xi = \sqrt{J_{eff}\over{h_{eff}}} \,.
\ee
The variation of $\xi$ with temperature, for fixed field is quite small
throughout the range of fields that interests us. In Figure 1 we show
 $(\xi(0)/\xi(T))^2$ as a function of temperature for three different
field strengths.
Even for $h/J = 0.5$, there is only a $10\%$ variation, up
to a temperature $T/J = 0.7$, above which the Hartree 
approximation breaks down. As shown below, the temperature dependence
of the resulting distribution function is even weaker than that for 
$\xi/L$ and for practical purposes it can be considered as temperature
independent.

\begin{figure}
\begin{center}
\epsfig{file=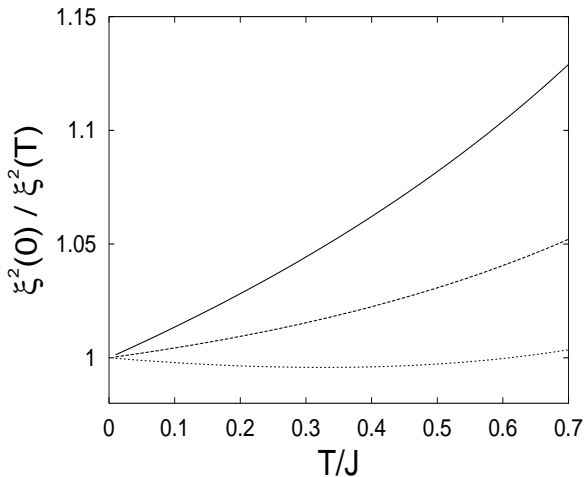,height=6.5cm,width=8cm}
\end{center}
\caption{$(\xi(0)/\xi(T))^2$, as given by the 
Hartree approximations for, from bottom to top, $h/J = 0.01, 0.05$ and $0.5$
(equations(\ref{H-1}) and (\ref{H-2})) as a function of $T/J$.}
\end{figure}
The parameters $h_{eff}$ and $\xi/L$ can be found in table 1
for a system of size $L=32$ at temperature $T/J = 0.7$ and for
field strength between $h=0.001$ and $h =0.5$. For finite field
the ratio $\xi/L$ varies from $\xi/L = 0.99$ to $\xi/L = 0.043$.

\begin{center}
\begin{tabular}[h]  {||l|l|l|l|l|l||}
\hline $h$ & $h_{eff}$ & $\xi/L$ & $a$ & $\gamma$  \\
\hline $0.0$ & $0.0$ & $\infty$ & $1.5807$  & $-0.89$   \\
\hline $0.001$ & $0.001$ & $0.987$ & $1.611$ & $-0.88$   \\
\hline $0.005$ & $0.005$ & $0.4407$ & $1.7416$  & $-0.844$   \\
\hline $0.01$  & $0.0101$ & $0.3111$ & $1.903$  & $-0.801$   \\
\hline $0.05$  & $0.0512$ & $0.1381$ & $3.359$ & $-0.583$   \\
\hline $0.1$  & $0.1034$ & $0.098$ & $5.333$ & $-0.463$   \\
\hline $0.5$ & $0.523$ & $0.0429$ & $21.82$ & $-0.23$   \\
\hline
\end{tabular}
\medskip

{\small TABLE 1 Variation of $h_{eff}$, $\xi/L$, $a$ and  $\gamma$
with field $h$ for $L=32$}
\end{center}

\section{The probability density function in finite field}

We have previously developed~\cite{pre,prl} the following expression for
the PDF
\be\label{eq-pdf}
P_L(\mu)&=&\int_{-\infty}^{\infty}\frac{dx}{2\pi}e^{ix\mu}\varphi(x) \\ \nonumber
\ln{\varphi(x)} & =&
-i x\sqrt{\frac{1}{2g_{2}}}
\sum_{{\bf q}\neq {\bf 0}} \frac{G({\bf q})}{N}    \\ \nonumber
&-&\frac{1}{2}\sum_{{\bf q}\neq {\bf 0}} 
 \ln\left[1-i\sqrt{\frac{2}{g_2}}{G({\bf q})\over{N}}x\right]
\ee
where $g_k = 1/N^k \sum_{\vec q} G(q)^k$  and $\mu = (m-\langle m
\rangle )/\sigma$.
The PDF with a finite correlation length is calculated from the same
expression by inserting the the modified Green's function (\ref{eq-G}).
The equivalence of equations (\ref{scale}) and
(\ref{eq-pdf}) is a result of the hyperscaling
relation~\cite{pre} and the functional
dependence predicted by the scaling
hypothesis (\ref{scale}) comes directly from
dimensional analysis of equation (\ref{eq-pdf}). We note that the
calculation can easily be extended to explore the non-Gaussian but
non-critical behaviour in all dimension less than four~\cite{pre}.
Summing over the Brillouin zone for a large but finite system and
performing a numerical Fourier transform we generate the
data shown in figures 2 and 3.
Data is shown in figure 2, for $h/J = 0.05$ ($\xi/L = 0.138$) and in Figure 3
for $h/J =0.01,$ $0.05$ and $0.5$ corresponding to $\xi/L=0.311,$ $0.138$
and $0.043$. It is compared
with results from  Monte Carlo simulation for a system of size $L=32$.
In each case, theoretical and numerical data are shown for three 
temperatures $T/J = 0.1, \; 0.4,\; 0.7$.
As in the zero field case,
 which
can be considered as the extreme non-Gaussian limit for such a system, the
PDFs are
characterized by an exponential tail for fluctuations below the mean
and a double exponential above the mean~\cite{pre}. Applying the field
reduces the asymmetry and in large field the data approach
a Gaussian distribution. 

Agreement between the theoretical calculation and the Monte Carlo simulation
is generally extremely good,
indicating that
the Hartree approximations are accurate. In Figure 2 all sets of
data collapse, within numerical error, onto a single curve independently 
of temperature.  This is the case for all field values chosen. 
When plotted on a logarithmic scale, temperature dependence is still not
observable, but a difference between theoretical
and numerical values can be observed
along the exponential tail, for probability densities
smaller than $10^{-4}$. The discrepancy appears largest
for fields around $h/J \sim 0.05$. This must indicate the limit of the
ability of the Hartree approximation in dealing with the fluctuations. 
For very small field, its effect is small and so errors are negligible, while
for larger fields critical fluctuations are smaller and one can imagine 
that the Hartree
approximation becomes quantitatively very accurate. Only in the intermediate
field range of $h/J \sim 0.05$, the combination of these two effects is 
sufficient to give the small deviation from the Monte Carlo
data.

\begin{figure}
\begin{center}
\epsfig{file=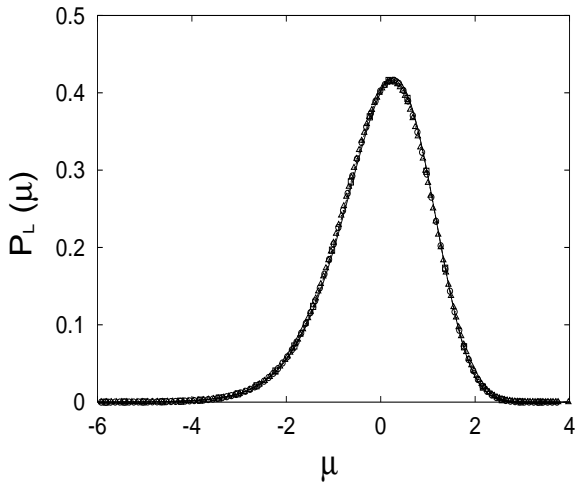,height=6.5cm,width=8cm}
\end{center}
\caption{Monte Carlo data  for 
$P_L(\mu)$ for
magnetic field  $h/J = 0.05$ and for $T/J = 0.1,0.4$ and $0.7$,
using the Hamiltonian (\ref{eq-H0}).
The lines are for
data generated from equation (\ref{eq-pdf}) (dotted) and for
the function (\ref{gumbel})
(dashed).}
\end{figure}

In all cases, both numerical and theoretical, the independence of the results
on temperature is quite remarkable and it leads one to suggest that, as
in the case of zero field, the distribution is truly temperature 
independent throughout the range of temperature and system size for which
the excitation of vortex pairs can be neglected. This point requires
further study, but it is already clear that, from a pragmatic point
of view, temperature dependence is not an observable phenomena.

Note that with the definition (\ref{eq-mag}) we explicitly study the longitudinal
magnetization, irrespective of its direction in space. This is the quantity 
that becomes critical at a phase transition in a system with continuous 
symmetry. The 
introduction of the magnetic field breaks the orientational symmetry in 
the thermodynamic limit and the variable conjugate to it is the projection
of $\vec m$ along the field direction. The fluctuations in these two quantities are different
for small field and
in the crossover region, as $h \rightarrow 0$,
the latter quantity becomes ill-defined~\cite{BHS}. The two quantities become 
indistinguishable
for $\xi/L \sim 0.1$.

The skewness
$\gamma = \langle \mu^3 \rangle $, which parameterizes the asymmetry of
the curve,
 varies from $-0.89$ in the
extreme non-Gaussian limit to $-0.23$ for a ratio $\xi/L =0.043$ and indeed
goes smoothly to zero for $\xi/L \rightarrow 0$.
Numerical values can also be found in
table 1.

\begin{figure}
\begin{center}
\epsfig{file=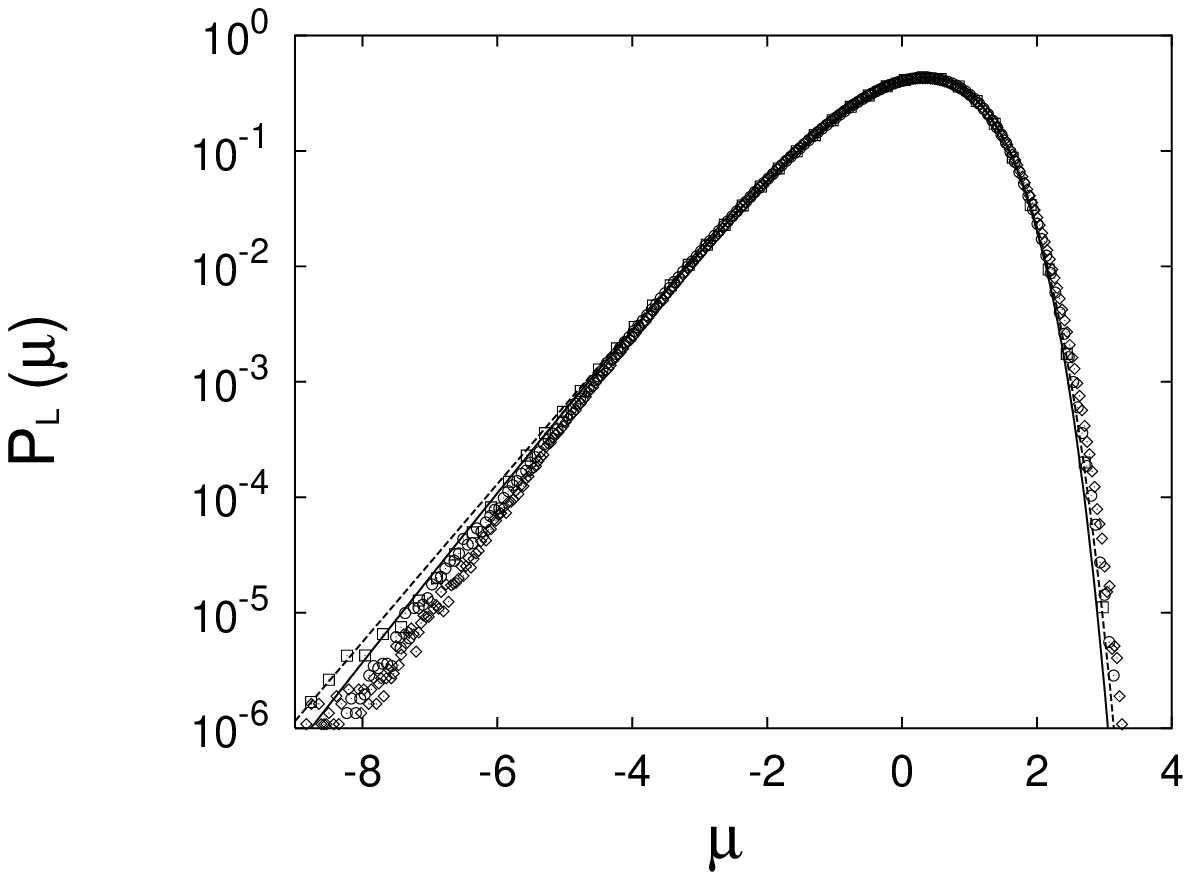,height=6.5cm,width=8cm}
\epsfig{file=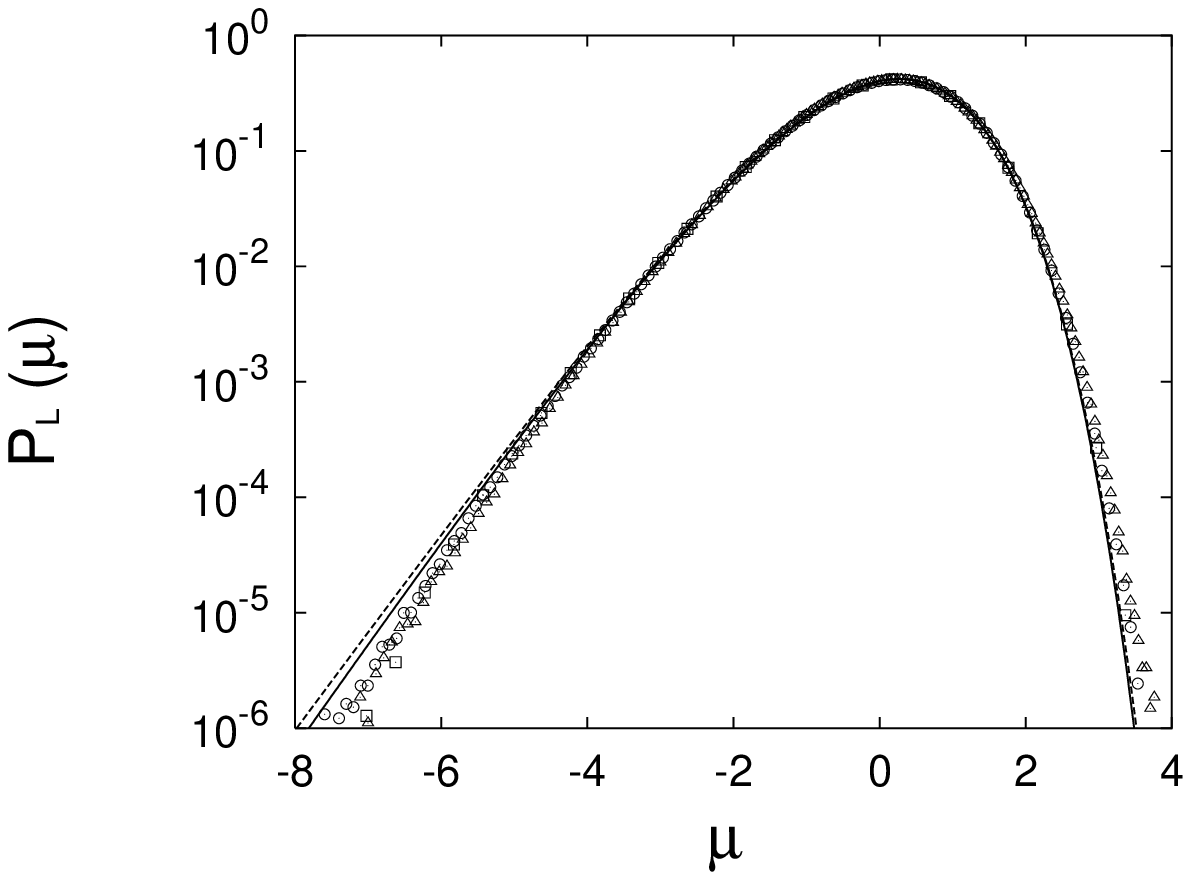,height=6.5cm,width=8cm}
\epsfig{file=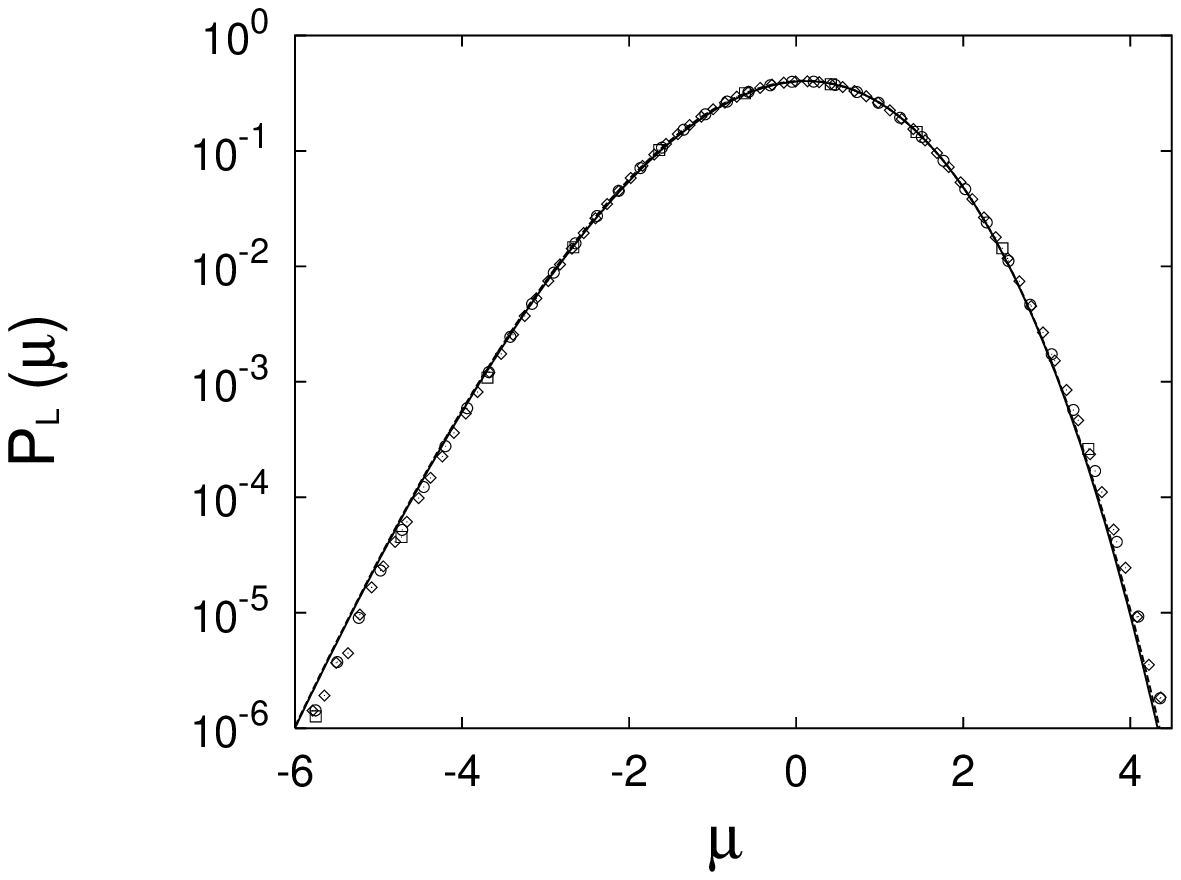,height=6.5cm,width=8cm}
\end{center}
\caption{Monte Carlo data for $P_L(\mu)$ for
magnetic field  (a) $h/J = 0.01$, (b) $h/J = 0.05$, (c) 
$h/J=0.5$ for $T/J = 0.1,0.4$ and $0.7$.
The lines are for
data generated from equation (\ref{eq-pdf}) (full) and for
the function (\ref{gumbel})
(dashed). Theoretical curves for different temperatures are
superposed.}
\end{figure}


The origin of the skewness becomes clear if one considers the
contribution made by the normal modes.
The magnetization can be written
\be
m = 1- 1/2N \sum_{q \rangle 0} \phi_q^2 + \dots \,.
\ee
To leading order in $\phi_q$ $m$ therefore consists of
a sum over variables $m_{\bf q} = (\phi_q^2/2N)$ which, within
the spin wave approximation, are statistically independent,
with generating function
\be\label{micro}
P(m_{\bf q}) = \sqrt{\beta J q^2 N\over{4 \pi}}
m_{\bf q}^{-1/2} \, {\rm e}^{-\beta J_{eff} N (q^2+\xi^{-2}) m_{\bf q}}\;\;.
\ee
In zero field the mean amplitudes $\langle m_{\bf q} \rangle $
vary from a microscopic value, $O(1/N)$ for modes on the zone edge through
to a value of $O(1)$ for the long wavelength modes
at the zone center and the dispersion in the contributions is
divergent in the thermodynamic limit.
In two-dimensions the density of states is linear in $q$ which
is just what is required to engage the entire zone~\cite{pre}.

Violation of the central limit theorem therefore arises because the individual
elements, although statistically independent are not individually
negligible. The modes of divergent amplitude
 near the Brillouin zone center are responsible for the
anisotropy although all parts of the zone are required for a
detailed reconstruction of $P(m)$.
Introduction of the length scale $\xi$ removes the divergence
for $q \rightarrow 0$ and re-establishes the criterion that the
statistically independent elements are individually negligible. In
the limit that $\xi/L \rightarrow 0$ the distribution becomes Gaussian.
If the thermodynamic limit is taken while keeping the ratio $\xi/L$
constant the amplitudes remains divergent, but the
contribution from the zone center becomes
progressively less, as the ratio $\xi/L$ is reduced and the skewness falls
to zero.

\section{Fitting with a generalized Gumbel function}

In refs~\cite{pre,prl} we have compared the functional
form of $P_L(\mu)$ with a series of standard expressions.
Although none are exact solutions, they all give good fits to the data
and therefore offer very useful analytical functions as well as giving some
insight into the physical processes responsible for the asymmetric PDF.
Here, we only pursue one of these, the generalized
Gumbel function for the statistics of extremes~\cite{gum}
\begin{eqnarray}\label{gumbel}
\sigma_z P_G(\mu_z) & =&  w {\rm e}^{a b(\mu_z-s)
                              -a {\rm e}^{b(\mu_z -s)}} \,,
\end{eqnarray}
which gives the $a^{th}$ largest or smallest values of a set
of $N$ random numbers, ${z_i}$, in the limit that $N \rightarrow \infty$.
For the smallest values,
 $\mu_z= (z-\langle  z  \rangle  )/\sigma_z$. The constants
$w$, $b$ and $s$ depend on $a$
through the three conditions of normalization,
$\langle \mu_z \rangle  = 0$ and $\langle \mu_z^2 \rangle  =1$
and one finds
\be
w & = & {a^a\alpha_a\over{\Gamma (a)}} \sigma_z \nonumber \\ 
b &= & 
\sqrt{ {1\over{\Gamma (a)}} {\partial^2 \Gamma (a)\over{\partial a^2}}- 
\left[ {1\over{\Gamma (a)}} {\partial \Gamma (a)\over{\partial a}}
\right]^2 } \nonumber \\
s & = &  {1\over{b}} 
\left[\log (a) - {1\over{\Gamma (a)}} {\partial \Gamma(a)\over{\partial a}} 
\right] \,.
\ee
The
function therefore has only
one parameter, which is calculated by 
comparing the Fourier
transform of (\ref{gumbel}) with $\Phi(x)$ of equation (\ref{eq-pdf}).
Using the notation of equation (\ref{eq-pdf}) we have, for the 
Gumbel function 
\be\label{dev-1}
\ln \varphi_G(x)&=&\ln\frac{w\Gamma(a)}{sa^a}-ix\left(s+\frac{\Psi(a)}{b}-\frac{\ln(a)}{b}\right)\\ \nonumber
&-&\frac{x^2}{2b^2}\Psi'(a)
+i\frac{x^3}{6b^3}\Psi''(a)\\ \nonumber
&+&\frac{x^4}{24b^4}\Psi'''(a)-i\frac{x^5}{120b^5}\Psi^{(4)}(a)+\cdots
\ee
where $\Psi(z)$ is the digamma function $\Gamma'(z)/\Gamma(z)$.
For $\Phi$ we have:
\be\label{dev-2}
\ln\varphi(x)=
-\frac{1}{2}x^2-i\frac{\sqrt{2}g_3}{3g_2^{3/2}}x^3 \\ \nonumber
+\frac{g_4}{2g_2^2}x^4
+i\frac{2\sqrt{2}g_5}{5g_2^{5/2}}x^5+\cdots ,
\ee
from which it follows that the constant $a$ is implicitly given by
\be\label{implicit}
{\Psi''(a)\over{\Psi'(a)^{3/2}}} = -2^{3/2} {g_3^{3/2}\over{g_2}} \,.
\ee 
For zero field the solution is $a\approx \pi/2$, rather than an
integer value, showing $P(m)$ is not simply an extreme value distribution.
That this solution is an approximation, though be it a good one, can be
seen by comparing the ratio of higher order terms in the two expansions
(\ref{dev-1}) and (\ref{dev-2}). These diverge slowly from unity~\cite{pre}.
Solving  equation (\ref{implicit}) for finite field gives $a$ 
increasing with $h$. The subsequent curves are superimposed in Figures 2 and 3
where one can see that the fitting function reproduces the results of 
the theoretical calculation to a good approximation. For small field 
a very small difference in the slopes of the exponential tails
can be detected. However, this disappears with increasing $h$ and the fitting
function can be regarded as an excellent working tool for describing the data.
In Figure 4 we illustrate the evolution of the distribution from the
anisotropic limit to the uncorrelated Gaussian limit as a function of field
using equation (\ref{gumbel}).
The values of $a(h)$ are shown
in table 1. In terms of extremal
statistics, the evolution of $a(h)$ means that we are describing the 
PDF of less and less extreme values, which
becomes more and more normal.

\begin{figure}
\begin{center}
\epsfig{file=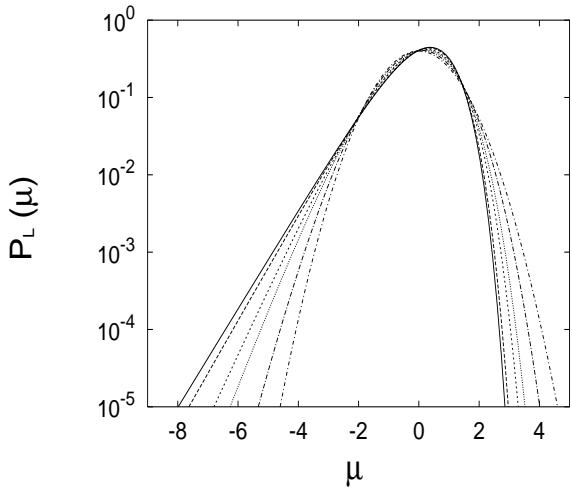,height=6.5cm,width=8cm}
\end{center}
\caption{Evolution of $P_L(\mu)$ from the function (\ref{gumbel})
with $h/J$. Moving from left to right for fluctuations below the mean,
the data is for $h/J=0, \; 0.01, \; 0.05,\;0.1,\; 0.5,\; \infty$}
\end{figure}

For strong field one can solve equation(\ref{implicit}) analytically.
 Evaluating $g_2$ and $g_3$ using a continuum approximation
and using Stirling's formula, $\ln \Gamma(a) \approx
a\ln a -a$, one finds
\begin{equation}
a \sim {\pi\over{2}} \left[1 + \left(\frac{L}{2\pi \xi} \right)^2 \right] \,.
\end{equation}
This simple expression gives $a$ to a good approximation even outside 
the range of $a$ values for which Sterling's formula is valid and 
reproduces our previous result for $h=0$. It also allows one to see 
that $a$ has a contribution
coming from fluctuations within a correlated domain and a contribution
coming from the fact that the system, with finite $\xi$, can be divided
into a number, $N_{eff} = (L/\xi)^2$, of statistically independent domains.

\section{conclusion}

In conclusion, we have made a microscopic calculation of the
generalized scaling function $P_L(mL^{\beta/\nu}, \xi/L)$ for order
parameter fluctuations near the line of critical points of the
low temperature phase of the two-dimensional XY model. 
A Hartree approximation is used to treat the non-linear 
corrections to a quadratic Hamiltonian. The approximation is 
necessary to ensure that the hyperscaling relation between
critical exponents is satisfied.
We show that the hyperscaling relation is a 
consequence of the non-linearity necessary
to ensure the correct system size dependence of the order parameter,
conjugate to the applied magnetic field. This is a requirement of
thermodynamics, rather than a general requirement for the
observation of non-Gaussian fluctuations for global quantities. Indeed,
observation of hyperscaling in non-thermodynamic systems~\cite{ban},
could be taken as an indication that an equivalent phenomenology exists.
For fixed magnetic field the correlation length is modified
only slightly by thermal fluctuations and this manifests itself in
the function $P_L$ which is essentially independent of temperature. 
The fact that the exponent $\beta/\nu = 4\pi T/J$ is
small and and that $\delta = 8\pi T/J -1$ is large may be important
for this observation. More work is required to clarify this point.

Finally,  we propose that our fitting parameter
$a(h)$ could be used as an
experimental tool to estimate the correlation length scale in 
other correlated systems.
We have previously made an empirical observation~\cite{BHP,prl} 
that the fluctuation
of global measures in other correlated systems, both in equilibrium
and out of equilibrium are very similar to those of the
magnetization of the 2D-XY model in zero field. We have proposed that
these observations illustrate at least qualitatively, 
universal features for correlated systems from different universality 
classes. This idea, as well as alternative 
interpretations~\cite{aji.01,fauve.01},
could be tested for example, in an
enclosed turbulent flow using the experimental
set up described in~\cite{LPF,PHL}, by varying the ratio of the power
injection length scale to the enclosure length scale.

It is a pleasure to thank our collaborators J.-Y. Fortin and J.-F. Pinton
for their contributions to this work and
L. Berthier, S. Fauve, A. Noullez and Z. R\'acz for useful discussions.
This work was supported by the PSMN at the 
\'Ecole Normale Sup\'erieure de Lyon.
M. S. is supported by E.U. contract ERBFMBICT983561.

\end{document}